\documentclass[twocolumn,showpacs,prl,superscriptaddress]{revtex4}          

\newcommand{ \be }{\begin{equation}}       
\newcommand{ \ee }{\end{equation}}       
\newcommand{ \bea }{\begin{eqnarray}}       
\newcommand{ \eea }{\end{eqnarray}}       
\newcommand{ \la }{\langle}       
\newcommand{ \ra }{\rangle}

\newcommand{ \mean }[1]{\left\langle #1 \right\rangle}   
\newcommand{ \etal }{{\it et al.}}   
\usepackage{color}

\usepackage{graphicx} 
\usepackage{dcolumn} 
\usepackage{bm} 

\begin{document}          
\title{       
\begin{flushright}  
{
\today \\  
 } 
\end{flushright} 
Directed flow in Au+Au collisions at $\sqrt{s_{_{NN}}} =$~62.4\,GeV    
} 
 
\affiliation{Argonne National Laboratory, Argonne, Illinois 60439}
\affiliation{University of Bern, 3012 Bern, Switzerland}
\affiliation{University of Birmingham, Birmingham, United Kingdom}
\affiliation{Brookhaven National Laboratory, Upton, New York 11973}
\affiliation{California Institute of Technology, Pasadena, California 91125}
\affiliation{University of California, Berkeley, California 94720}
\affiliation{University of California, Davis, California 95616}
\affiliation{University of California, Los Angeles, California 90095}
\affiliation{Carnegie Mellon University, Pittsburgh, Pennsylvania 15213}
\affiliation{Creighton University, Omaha, Nebraska 68178}
\affiliation{Nuclear Physics Institute AS CR, 250 68 \v{R}e\v{z}/Prague, Czech Republic}
\affiliation{Laboratory for High Energy (JINR), Dubna, Russia}
\affiliation{Particle Physics Laboratory (JINR), Dubna, Russia}
\affiliation{University of Frankfurt, Frankfurt, Germany}
\affiliation{Institute of Physics, Bhubaneswar 751005, India}
\affiliation{Indian Institute of Technology, Mumbai, India}
\affiliation{Indiana University, Bloomington, Indiana 47408}
\affiliation{Institut de Recherches Subatomiques, Strasbourg, France}
\affiliation{University of Jammu, Jammu 180001, India}
\affiliation{Kent State University, Kent, Ohio 44242}
\affiliation{Lawrence Berkeley National Laboratory, Berkeley, California 94720}
\affiliation{Massachusetts Institute of Technology, Cambridge, MA 02139-4307}
\affiliation{Max-Planck-Institut f\"ur Physik, Munich, Germany}
\affiliation{Michigan State University, East Lansing, Michigan 48824}
\affiliation{Moscow Engineering Physics Institute, Moscow Russia}
\affiliation{City College of New York, New York City, New York 10031}
\affiliation{NIKHEF and Utrecht University, Amsterdam, The Netherlands}
\affiliation{Ohio State University, Columbus, Ohio 43210}
\affiliation{Panjab University, Chandigarh 160014, India}
\affiliation{Pennsylvania State University, University Park, Pennsylvania 16802}
\affiliation{Institute of High Energy Physics, Protvino, Russia}
\affiliation{Purdue University, West Lafayette, Indiana 47907}
\affiliation{Pusan National University, Pusan, Republic of Korea}
\affiliation{University of Rajasthan, Jaipur 302004, India}
\affiliation{Rice University, Houston, Texas 77251}
\affiliation{Universidade de Sao Paulo, Sao Paulo, Brazil}
\affiliation{University of Science \& Technology of China, Anhui 230027, China}
\affiliation{Shanghai Institute of Applied Physics, Shanghai 201800, China}
\affiliation{SUBATECH, Nantes, France}
\affiliation{Texas A\&M University, College Station, Texas 77843}
\affiliation{University of Texas, Austin, Texas 78712}
\affiliation{Tsinghua University, Beijing 100084, China}
\affiliation{Valparaiso University, Valparaiso, Indiana 46383}
\affiliation{Variable Energy Cyclotron Centre, Kolkata 700064, India}
\affiliation{Warsaw University of Technology, Warsaw, Poland}
\affiliation{University of Washington, Seattle, Washington 98195}
\affiliation{Wayne State University, Detroit, Michigan 48201}
\affiliation{Institute of Particle Physics, CCNU (HZNU), Wuhan 430079, China}
\affiliation{Yale University, New Haven, Connecticut 06520}
\affiliation{University of Zagreb, Zagreb, HR-10002, Croatia}

\author{J.~Adams}\affiliation{University of Birmingham, Birmingham, United Kingdom}
\author{M.M.~Aggarwal}\affiliation{Panjab University, Chandigarh 160014, India}
\author{Z.~Ahammed}\affiliation{Variable Energy Cyclotron Centre, Kolkata 700064, India}
\author{J.~Amonett}\affiliation{Kent State University, Kent, Ohio 44242}
\author{B.D.~Anderson}\affiliation{Kent State University, Kent, Ohio 44242}
\author{D.~Arkhipkin}\affiliation{Particle Physics Laboratory (JINR), Dubna, Russia}
\author{G.S.~Averichev}\affiliation{Laboratory for High Energy (JINR), Dubna, Russia}
\author{S.K.~Badyal}\affiliation{University of Jammu, Jammu 180001, India}
\author{Y.~Bai}\affiliation{NIKHEF and Utrecht University, Amsterdam, The Netherlands}
\author{J.~Balewski}\affiliation{Indiana University, Bloomington, Indiana 47408}
% ...

\collaboration{STAR Collaboration}\noaffiliation
   
 
\begin{abstract}     
We present the directed flow ($v_1$) measured in Au+Au collisions at
$\sqrt{s_{_{NN}}} = 62.4$\,GeV in the mid-pseudorapidity region
$|\eta|<1.3$ and in the forward pseudorapidity region
$2.5 < |\eta| < 4.0$.  The results are obtained using
the three-particle cumulant method, the event plane method with
mixed harmonics, and for the first time at the Relativistic Heavy Ion
Collider (RHIC), the standard method with the
event plane reconstructed from spectator neutrons.
Results from all three methods are in good agreement.
Over the pseudorapidity range studied,
charged particle directed flow is in the direction
opposite to that of fragmentation neutrons.
\end{abstract}
 
  
\maketitle

Directed flow in heavy-ion collisions is quantified by the
first harmonic ($v_1$) in the Fourier expansion of the azimuthal
distribution of produced particles with respect to the reaction
plane~\cite{Methods}.  It describes collective sideward motion of
produced particles and nuclear fragments and carries information on
the very early stages of the collision~\cite{sorge}.  The shape of
$v_1(y)$ in the central rapidity region is of special interest because
it might reveal a signature of a possible Quark-Gluon Plasma (QGP)
phase~\cite{antiflow,third-component,Horst}.
  
At AGS and SPS energies, $v_1$ versus rapidity is an almost linear function of
rapidity~\cite{v1LowerEnergies,e877,na49}.
Often, just the slope of $v_1(y)$ at midrapidity is used to define
the strength of directed flow.  The sign of $v_1$ is by convention
defined as positive for nucleons in the projectile fragmentation
region.  At these energies,
the slope of $v_1(y)$ at midrapidity is observed to be
positive for protons, and significantly smaller in magnitude and
negative for pions~\cite{e877,na49,wa98}.  The opposite directed
flow of pions is usually explained in terms of shadowing by nucleons.
At RHIC energies, directed flow is predicted to be smaller
near midrapidity with a weaker dependence on pseudorapidity
\cite{wiggle,bleicher}.  It may exhibit a characteristic
wiggle~\cite{antiflow,third-component,wiggle,bleicher}, whereby
directed flow changes sign three times outside the beam
fragmentation regions, in contrast to the observed sideward
deflection pattern at lower energies where the sign of $v_1(y)$
changes only once, at midrapidity.
The observation of the slope of $v_1$
at midrapidity being negative for
nucleons or positive for pions would constitute such a wiggle.

In one-fluid dynamical calculations~\cite{antiflow,third-component}, 
the wiggle structure appears only under the assumption of a QGP 
equation of state, thus becoming a signature of the QGP phase  
transition.  Then the wiggle structure  is interpreted to be a  
consequence of the expansion of the highly compressed, disk-shaped  
system, with the plane of the disk initially tilted with respect to  
the beam direction.~\cite{antiflow}  The subsequent system expansion leads to the 
so-called anti-flow~\cite{antiflow} or third flow 
component~\cite{third-component}.  
Such flow can reverse the normal pattern of sideward 
deflection as seen at lower energies, and hence can result in either a 
flatness of $v_1$, or a wiggle structure if the expansion is 
strong enough.  A similar wiggle structure in nucleon $v_1(y)$ is 
predicted if one assumes strong but incomplete baryon stopping 
together with strong space-momentum correlations caused by transverse 
radial expansion~\cite{wiggle}.  While the predictions for baryon 
directed flow are unambiguous in both hydrodynamical and  
transport models, the situation for pion directed flow is less clear.   
RQMD model calculations~\cite{wiggle} for Au+Au collisions  
at $\sqrt{s_{NN}} = 200$\,GeV indicate that shadowing by protons  
causes the pions to flow mostly with opposite sign to the protons,  
but somewhat diffused due to higher thermal velocities for pions.  
Similar UrQMD calculations~\cite{bleicher} predict no wiggle for 
pions in the central rapidity region with a negative slope at 
midrapidity as observed at lower collision energies.

At RHIC, most of the detectors 
cover the central rapidity region where the directed flow signal is 
small and the analysis procedures easily can be confused by azimuthal 
correlations not related to the reaction plane orientation, the 
so-called non-flow effects. Only recently have the first $v_1$ results been 
reported by the STAR Collaboration~\cite{v1v4} and preliminary results by the 
PHOBOS Collaboration~\cite{PhobosQM2004}.   
In ~\cite{v1v4}, the shape of $v_1$ in the 
region on either side of midrapidity is poorly resolved due to large 
statistical errors.  This shortcoming arose from having only about 
70,000 events from the Forward Time Projection Chambers (FTPCs)~\cite{FTPC-NIM} during 
their commissioning in the RHIC run II period (2002). 

In this paper, we present directed flow measurements
in Au+Au collisions at $\sqrt{s_{NN}} =62.4$\,GeV.
Results are obtained by three different methods, namely,
the three-particle cumulant method ($v_1\{3\}$)~\cite{Borghini}, the event plane
method with mixed harmonics ($v_1\{\mathrm{EP}_1,\mathrm{EP}_2\}$)~\cite{Methods,Flow200GeV},
and the standard method~\cite{Methods} with the first-order
event plane reconstructed from neutral fragments of the incident
beams ($v_1\{\text{ZDC-SMD}\}$). 
According to the standard method~\cite{Methods}, directed flow can be evaluated by
\begin{equation}
 v_1\{\text{Standard}\}=\langle \cos(\phi - \Psi_1) \rangle /\text{Res}(\Psi_1)
\label{equ:Standard}
\end{equation}
where $\phi$ and $\Psi_1$ denote the azimuthal angle of the particle and the first-order
event plane, respectively, and Res($\Psi_1)=\mean{\cos(\Psi_1-\Psi_{\text{RP}})}$ represents the resolution of
the first-order event plane.
In the three-particle cumulant method~\cite{Borghini},
the flow is evaluated from
\begin{eqnarray}
\langle\langle\cos (\phi_a+\phi_b-2\phi_c)\rangle\rangle \equiv \langle\cos(\phi_a+\phi_b-2\phi_c)\rangle \nonumber \\
- \langle\cos(\phi_a+\phi_b)\rangle\langle\cos(-2\phi_c)\rangle \nonumber \\
- \langle\cos\phi_a\rangle\langle\cos(\phi_b-2\phi_c)\rangle \nonumber \\
- \langle\cos\phi_b\rangle\langle\cos(\phi_a-2\phi_c)\rangle \nonumber \\
+ 2\langle\cos\phi_a\rangle\langle\cos\phi_b\rangle\langle\cos(-2\phi_c)\rangle \nonumber \\
=v_{1,a}v_{1,b}v_{2,c}        ~~~~~~~~~~~~~
\label{equ:Cumulant} 
\end{eqnarray}
where on the r.h.s.\ of the first equality, the first term is a three-particle correlation and the
other terms are to isolate the genuine three-particle correlation from 
spurious correlations induced by detector effects. Subscripts $a$, $b$ and $c$ 
denote three different particles. 
This method was used in the
first $v_1$ publication at RHIC~\cite{v1v4}.
The event plane method with mixed harmonics~\cite{Flow200GeV} utilizes 
the second-order event plane from the TPC, $\Psi_2^{\mathrm{TPC}}$, and two first-order
event planes from random subevents in the FTPCs, 
$\Psi^{\mathrm{FTPC}_1}_1$ and $\Psi^{\mathrm{FTPC}_2}_1$. It measures
\begin{eqnarray} \label{eq:v1ep1ep2}
v_1\{\mathrm{EP}_1,\mathrm{EP}_2\} =
\end{eqnarray}
\begin{eqnarray}
\frac{\left\la\cos\left(\phi+\Psi_1^{\mathrm{FTPC}}-2\Psi_2^{\mathrm{TPC}}\right)\right\ra}{\sqrt{\left\la\cos\left(\Psi_1^{\mathrm{FTPC}_1}+\Psi_1^{\mathrm{FTPC}_2}-2\Psi_2^{\mathrm{TPC}}\right)\right\ra\cdot \mathrm{Res}(\Psi_2^{\mathrm{TPC}})}}\nonumber
\end{eqnarray}
where the emission angle of the particle $\phi$ is correlated with the
first-order event plane $\Psi_1^{\mathrm{FTPC}}$ of the 
random subevent (made of tracks of both FTPCs) which does not contain the particle. 
Res($\Psi_2^{\mathrm{TPC}}$) represents the resolution of the second-order event plane measured in
the TPC~\cite{Flow200GeV}.
Both the cumulant method and the event plane method with mixed harmonics 
offer enhanced suppression of non-flow
effects, including correlations due to momentum conservation,
compared with the standard method (in which the event plane is
reconstructed for the same harmonics and in the same rapidity
region where the event anisotropy is measured). In the
present study, the procedures to obtain $v_1\{3\}$ and
$v_1\{\mathrm{EP}_1,\mathrm{EP}_2\}$ are essentially the same as in
Ref.~\cite{v1v4}.  
\begin{table}[hbt]
\begin{center}
\begin{tabular}{c|c}\hline \hline
Centrality &  Event plane resolution
\\ \hline
$70\%-80\%$  & $0.179\pm 0.005$
\\ \hline
     $60\%-70\%$  & $0.185\pm 0.004$
\\ \hline
     $50\%-60\%$  & $0.176\pm 0.005$
\\ \hline
 $40\%-50\%$       & $0.167\pm 0.005$
\\ \hline
 $30\%-40\%$     & $0.138\pm 0.006$
\\ \hline
  $20\%-30\%$  & $0.110\pm 0.008$
\\ \hline
 $10\%-20\%$& $0.081\pm 0.010$
\\    \hline \hline
\end{tabular}
\end{center}
\caption{The resolution of the first-order event plane~\cite{Methods} provided by the
ZDC-SMDs, as determined from the sub-event correlation between east and
west SMDs. The errors in the table are statistical.
}
\label{tbl:resolution}
\end{table}
\begin{figure}[t]
  \includegraphics[width=0.50\textwidth]{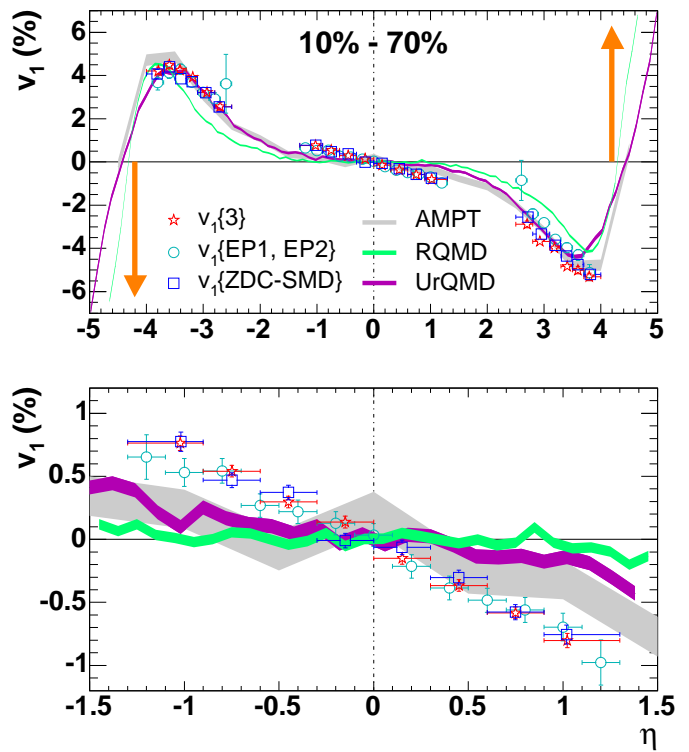}
  \caption{(color online) Directed flow of charged particles
    as a function of pseudorapidity, for centrality 10\%--70\%.  The
    arrows in the upper panel indicate the direction of flow
    for spectator neutrons.  The arrow positions on the pseudorapidity
    axis corresponds to where the incident ions would lie on a
    rapidity scale.  The lower panel shows the mid-pseudorapidity
    region in more detail. The plotted errors are statistical only, and systematic 
    effects are discussed in the text.}
\label{fig:OneCentBin3Methods}
\end{figure}
In the third method, the reaction plane was
determined from the sideward deflection of spectator neutrons
("bounce-off")~\cite{v1LowerEnergies} measured in the Zero Degree
Calorimeters (ZDCs)~\cite{ZDC}.
This is the first report 
from RHIC of flow results with the event plane reconstructed from 
spectator fragments. Five million minimum-bias events were 
used in this study for each of the three analyses, and all the  
errors presented are statistical. Cuts used in the 
TPC ($|\eta| < 1.3$)~\cite{TPC-NIM} and FTPC ($2.5 < |\eta| < 4.0$) analyses  
are the same as listed in Table II of Ref.~\cite{Flow200GeV},  
except that the vertex $z$ cut is from $-30$ to 30 cm. 
The centrality definition, based on the
raw charged particle TPC multiplicity with $|\eta| < 0.5$, is
the same as used previously~\cite{Flow200GeV}.
 
In the Fall of 2003, STAR installed Shower Maximum Detectors (SMDs)
sandwiched between the first and second modules of each existing STAR
ZDC at $|\eta| > 6.3$.  Each SMD consists of two plastic scintillator planes, one of 7
vertical slats and another of 8 horizontal slats.  The two SMDs
provide event-by-event information on the transverse distribution of
energy deposition of the spectator neutrons. The weighted
center of the energy distribution determines the event plane vector on
each side. The combination of the east and west event plane vectors
provides the full event plane and the event plane resolution is
obtained from the correlation of the two event plane vectors in the
standard way.
The $v_1\{\text{ZDC-SMD}\}$ should have minimal contribution from
non-flow effects due to the large rapidity gap between the spectator
neutrons used to establish the reaction plane and the rapidity region
where the measurements were performed.  The resolution, as
defined in~\cite{Methods}, of the first-order event plane
reconstructed using the ZDC-SMDs is listed in
Table~\ref{tbl:resolution}.
\begin{figure}[t]
  \includegraphics[width=0.50\textwidth]{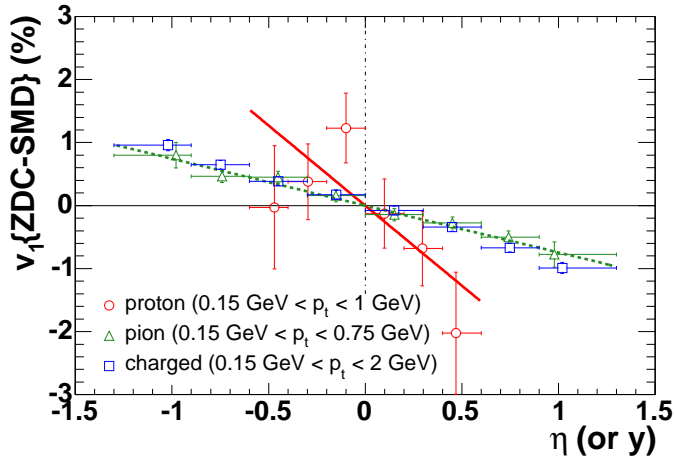}
  \caption{
    (color online) $v_1$ versus rapidity for protons and pions. The
    charged particle $v_1 (\eta)$ is plotted as a reference. The
    different upper end of the $p_t$ range for protons and pions is
    due to different limits of the $dE/dx$ identification method. The
    solid and dashed lines are results from linear fits described in
    the text. All results are from analyses using the reaction plane
    reconstructed by the ZDC-SMD, $v_1\{\text{ZDC-SMD}\}$.
    The plotted errors are statistical only, and systematic 
    effects are discussed in the text.}
\label{fig:PidV1_62GeV40Pct70Pct}
\end{figure}
\begin{figure*}[t]
\vspace{-0.55cm}
  \includegraphics[width=1.0\textwidth]{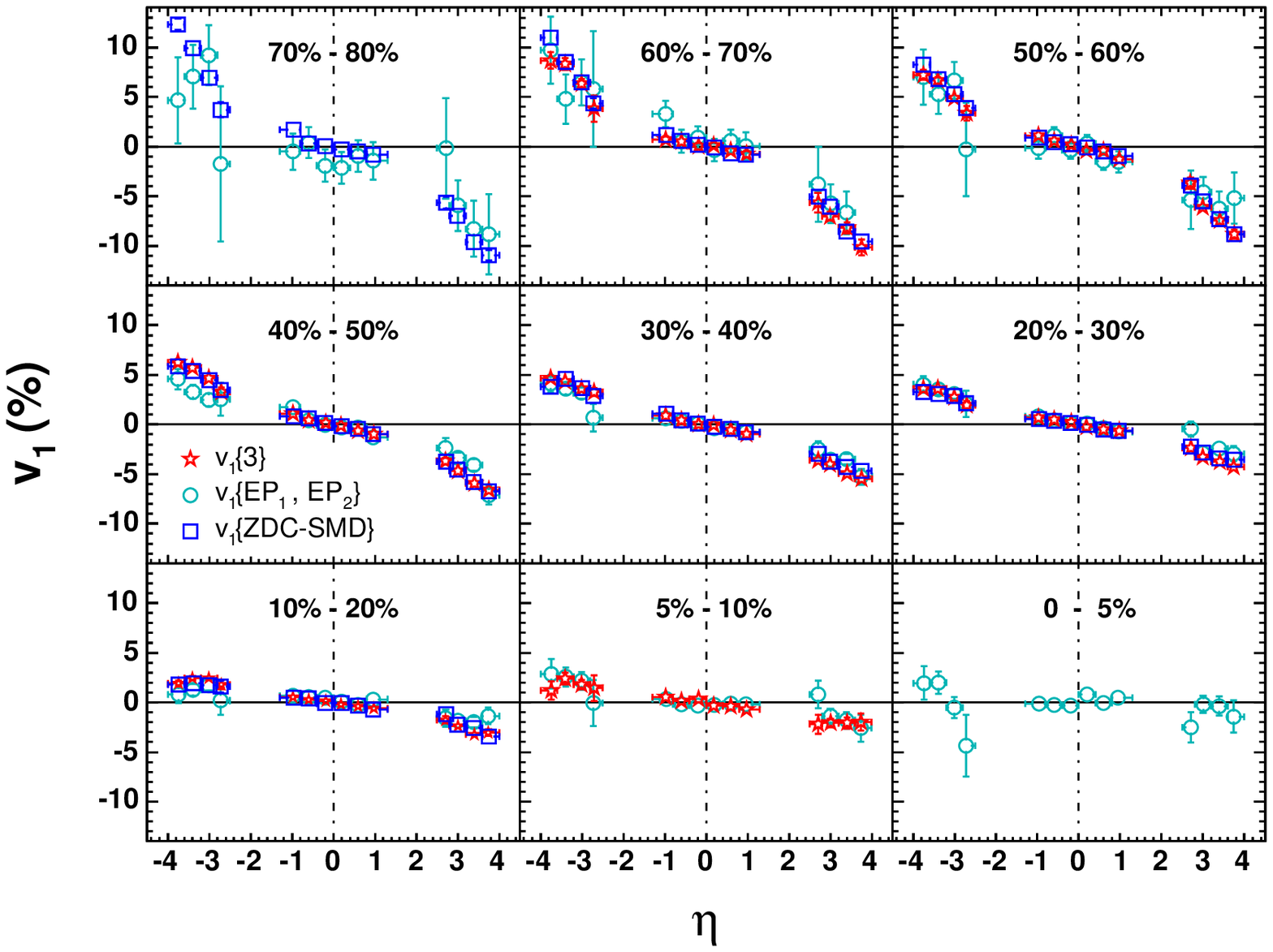}
\vspace{-1cm}
  \caption{(color online) Directed flow of charged particles
as a function of pseudorapidity for different centralities.
The plotted errors are statistical. }
  \label{fig:AllCent3Methods}
\end{figure*}
 
The centrality ranges of Au + Au collisions at $\sqrt{s_{NN}} =
62.4$\,GeV where the three $v_1$ methods work are slightly
different:  $v_1\{3\}$ fails at centralities less than 5\% and
centralities greater than 70\%, because the four particle cumulant
$v_2\{4\}$, which is a necessary ingredient for measuring $v_1\{3\}$,
is not measurable in those regions possibly due to large $v_2$
fluctuations; $v_1\{\text{ZDC-SMD}\}$ fails for centrality less than
10\% because of insufficient event plane resolution in central
collisions.  Figure~\ref{fig:OneCentBin3Methods} shows charged particle
$v_1$ as a function of pseudorapidity, $\eta$, for centrality
10\%--70\% where all three methods work, from Au + Au collisions at
$\sqrt{s_{NN}} = 62.4$\,GeV.  The arrows in the upper panel
indicate the direction of flow for spectator neutrons as determined
from the ZDC-SMDs.  The lower panel shows on expanded scales
the mid-pseudorapidity region measured by the STAR TPC.  The results
from the three different methods agree with each other very well.  In
Ref.~\cite{v1v4}, the relative systematic uncertainty in $v_1\{3\}$
and $v_1\{\mathrm{EP}_1,\mathrm{EP}_2\}$ was estimated to be about
20\%.  That error estimate was obtained under the assumption that the
directed flow measurements using two-particle correlations were
totally dominated by non-flow effects.  Such an assumption provides an
upper limit on the systematic errors.
Ref.~\cite{Flow200GeV} provides further discussion on the
systematic uncertainties.
The comparison of $v_1\{\text{ZDC-SMD}\}$ and $v_1\{3\}$ indeed shows
that the relative difference is no more than 20\% around mid-pseudorapidity
(where the directed flow itself is less than 0.005) and the difference
is only about 5\% in the forward pseudorapidity region.
$v_1\{\text{ZDC-SMD}\}$ was also calculated using the information from
the east and west ZDCs separately as well as separately from
correlations in the vertical and horizontal directions (note that
the ZDC-SMDs have a rectangular shape); all the results
agree within 15\%.
In another systematic study of $v_{1}\{$ZDC-SMD$\}$, a tighter distance of the closest approach (dca) cut was applied to reduce the number of weak decay tracks or secondary interactions. 
The ratio of $v_1$ obtained with dca $<1$ cm to the $v_1$ result with the default cut (dca $<2$ cm) was measured to be 
$v_1^\mathrm{dca<1\,cm}/v_1^\mathrm{dca<2\,cm} = 1.00 \pm 0.07$ for charged particles.
 
AMPT~\cite{ampt}, RQMD~\cite{sorge}, and UrQMD~\cite{urqmd} model
calculations for the same centrality of Au + Au collisions at
$\sqrt{s_{NN}} = 62.4$\,GeV are also shown in
Fig.~\ref{fig:OneCentBin3Methods}.  Most transport models, including
AMPT, RQMD and UrQMD, underpredict elliptic flow ($v_2$) at RHIC 
energies, and we now report that they also underpredict the 
charged particle $v_1(\eta)$ within a unit or so of 
mid-pseudorapidity, but then come into good agreement with 
the data over the region $2.5<|\eta|<4.0$.  While 
the magnitude of $v_1$ for charged particles increases with 
the magnitude of pseudorapidity below $|\eta| \sim 3.8$ for 
centralities between 10\% and 70\%, our results are compatible 
with the peak in $|v_1|$ lying in the $|\eta|$ region predicted 
by all three models, namely, 3.5 to 4.0. 
\begin{figure}[t]
  \includegraphics[width=0.45\textwidth]{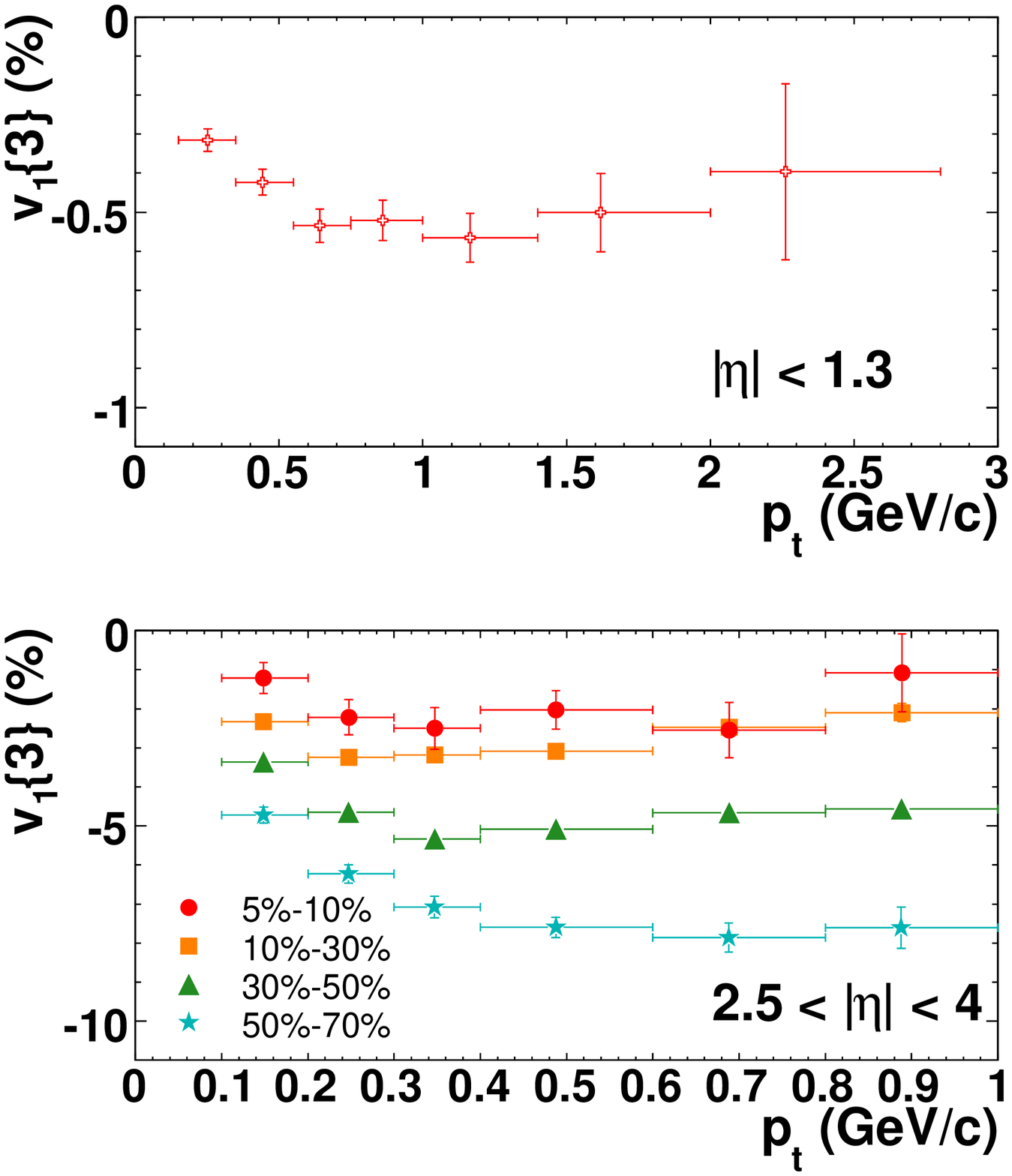}
  \caption{(color online)
  The upper panel shows $v_1\{3\}$ versus $p_t$ measured in the main
  TPC ($|\eta|<1.3$), for centrality 10\%--70\%.  The lower panel shows
  $v_1\{3\}$ versus $p_t$ measured by the Forward TPC ($2.5<|\eta|<4.0$), for
  different centralities. Note the different scales on both axes for
  the two panels. 
  The differential directed flow of particles with negative $\eta$
  has been changed in sign as stated in the text.
  The plotted errors are statistical.}  \label{fig:v1Pt}
\end{figure}

No apparent wiggle structure, as discussed above, is observed 
within our acceptance.
Throughout our pseudorapidity acceptance, charged particles on a given
side of $\eta =0$ flow in the opposite direction to the fragmentation
neutrons on that side. This is consistent with the direction expected
in the ``anti-flow'' scenario~\cite{antiflow} but it is also the same
direction as measured for pions at lower energies that is usually
related to the pion shadowing by nucleons.  Assuming that the charged
particle flow at beam rapidity is dominated by protons, one would
conclude that over the entire pseudorapidity range $v_1(\eta)$
changes sign three times.  However, this does not prove the existence
of the wiggle structure for protons and pions separately.
Measurements of directed flow of identified particles could be more
informative in this respect.  In STAR, particle identification is
feasible only in the main TPC, which covers the pseudorapidity region
$|\eta| < 1.3$.  In this region, the RQMD model predicts very flat
$v_1(\eta)$ for pions and a clear wiggle structure, with negative
slope $d v_1/d \eta$ at mid-pseudorapidity for protons at $\sqrt{s_{NN}} =
62.4$\,GeV.  (The relatively strong wiggle for pions reported in
Ref.~\cite{wiggle} is developed only at higher collision energies.)
To maximize the magnitude of the possible slope, we select the
centrality interval 40\% to 70\%, where flow anisotropies normally are
at their peak. The result is shown in
Fig.~\ref{fig:PidV1_62GeV40Pct70Pct}.  With the current statistics, we
observe that pion flow is very similar to that of charged particles,
with the slope at midrapidity $d v_1/dy$ about $-0.0074\pm 0.0010$,
obtained from a linear fit over the region $|y| < 1.3$ (dashed line).
For protons, the slope $d v_1/dy$ is $-0.025 \pm 0.011$ from a
linear fit in $|y| < 0.6$ (solid line).  At present, STAR's statistics
for baryons are rather small compared with the statistics for all
charged particles, and our best estimates of the fitted slope are such
that a negative baryon slope with comparable magnitude to the RQMD
prediction is not decisively ruled out.
For the identified particles, the influence of the particle identification procedures on the flow values for pions and protons may be a source of errors.
By default we eliminate particles $3\sigma $ away from the expected TPC energy loss for the relevant particle type. 
When we tightened the cut to $2\sigma $ instead of $3\sigma $, we found that for 40\%--70\% most central events, 
the $v_1\{\text{ZDC-SMD}\}$ of pions is reduced by less than 10\% while the proton 
$v_1\{\text{ZDC-SMD}\}$ stays constant within errors.

Figure~\ref{fig:AllCent3Methods} shows $v_1$ of charged particles as a
function of $\eta$ for different centralities.
We do not observe an onset of any special feature in the 
pseudorapidity dependence of $v_1$ at any centrality.
Preliminary $v_1(\eta)$ results from PHOBOS \cite{PhobosQM05} 
for centrality 0 to 40\% are consistent with our data at the 
same centrality except that $|v_1(\eta)|$ from PHOBOS has 
its peak at $|\eta|$ of about 3 to 3.5, while STAR's 
$|v_1(\eta)|$ peaks at $|\eta|$ about 3.8 or higher. A
significant change in particle abundances below STAR's
transverse momentum acceptance cut (0.15 GeV$/c$) might
account for some or all of this difference in the $|v_1|$
peak position.

The transverse-momentum dependence of $v_1$ is shown in
Fig.~\ref{fig:v1Pt}.
Since $v_1(\eta, p_t)$ is asymmetric about $\eta=0$,
the integral of $v_1(\eta, p_t)$ over a symmetric $\eta$ range goes to zero.
We change $v_1(\eta, p_t)$ of particles with negative $\eta$
into $-v_1(-\eta, p_t)$, and integrate over all $\eta$.
Due to the small magnitude of the $v_1$ signal
close to mid-pseudorapidity ($|\eta|<1.3$), only the averaged $v_1(p_t)$
over centralities 10\%--70\% is shown.  For $2.5<|\eta|<4.0$, the
$v_1$ signal is large enough to be resolved for different centrality
regions.  
The poor $p_t$ resolution for higher $p_t$ in FTPCs limits the
$p_t$ range to below 1 GeV$/c$ for $2.5<|\eta|<4.0$.
For all centralities and pseudorapidity regions,
the magnitude of $v_1$ is observed to 
reach its maximum at $p_t \approx 1$ GeV$/c$ for $|\eta|< 1.3$ and
at $p_t \approx 0.5$ GeV$/c$ for $2.5<|\eta|<4.0$.
\begin{table}[hbt]
\begin{center}
\begin{tabular}{c|c}\hline \hline
Centrality &  Impact parameter (fm)
\\ \hline
$70\%-80\%$  & $12.82 + 0.62 - 0.67$
\\ \hline
     $60\%-70\%$  & $11.89 + 0.67 - 0.52$
\\ \hline
     $50\%-60\%$  & $10.95 + 0.58 - 0.52$
\\ \hline
 $40\%-50\%$       & $9.91 + 0.47 - 0.42$
\\ \hline
 $30\%-40\%$     & $8.71 + 0.52 - 0.31$
\\ \hline
  $20\%-30\%$  & $7.36 + 0.47 - 0.26$
\\ \hline
 $10\%-20\%$& $5.72 + 0.32 - 0.21$
\\ \hline
 $5\%-10\%$& $4.08 + 0.16 - 0.21$
\\ \hline
 $0-5\%$& $2.24 + 0.07 - 0.14$
\\    \hline \hline
\end{tabular}
\end{center}
\caption{The correspondence between centrality and impact parameter.  }
\label{tbl:impctPara}
\end{table}
\begin{figure}[t]
  \includegraphics[width=0.45\textwidth]{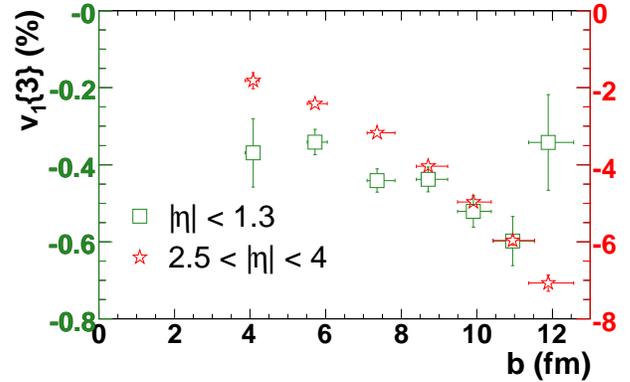}
  \caption{(color online)
  Directed flow of charged particles as a function of impact
  parameter for the mid-pseudorapidity region ($|\eta|< 1.3$, with the left vertical scale) and the forward
  pseudorapidity region ($2.5 <|\eta|< 4.0$, with the right vertical scale).
  The differential directed flow of particles with negative $\eta$
  has been changed in sign as stated in the text.
  The plotted errors are statistical.  }  
  \label{fig:v1Cent}
\end{figure}
Note that from its definition, $v_1(p_t)$ must approach zero as $p_t$ approaches zero. 
The centrality dependence of $p_t$-integrated $v_1$ is shown in 
Fig.~\ref{fig:v1Cent}. 
The values of the impact parameter were obtained using a Monte Carlo Glauber calculation \cite{Glauber}, 
listed in Table~\ref{tbl:impctPara}.
As expected, $v_1$ decreases with 
centrality.  It is seen that $v_1$ in the more forward 
pseudorapidity region $2.5<|\eta|<4.0$ varies more strongly with 
centrality than in the region closer to mid-pseudorapidity 
($|\eta|<1.3$). 
  
\begin{figure}[t]         
  \includegraphics[width=0.50\textwidth]{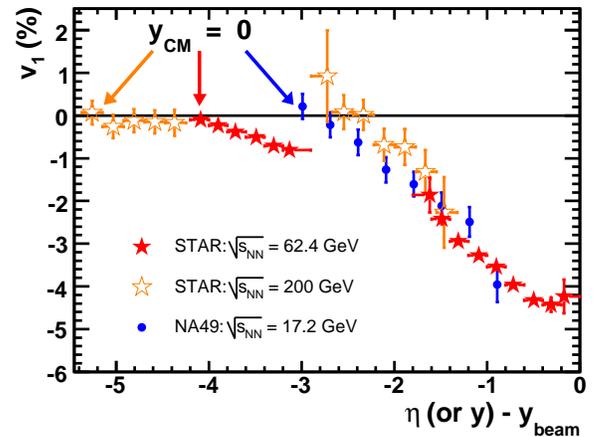}      
  \caption{  
    (color online) Charged particle $v_1$ for Au+Au collisions (10\%--70\%) at
    200~GeV~\cite{v1v4} (open stars) and 62.4~GeV (solid stars), as a function of
    $\eta - y_\mathrm{beam}$.  Also shown are results from NA49~\cite{na49}
    (circles) for pions from 158$A$ GeV midcentral (12.5\%--33.5\%)
    Pb+Pb collisions as a function of $y - y_\mathrm{beam}$.  The
    62.4~GeV and 200~GeV points are averaged over the positive and
    negative rapidity regions. All results are from analyses involving
    three-particle cumulants, $v_1\{3\}$. The plotted errors are statistical.}
  \label{fig:v1Eta3Energies}  
\end{figure}

It has been observed that particle emission (both spectra and flow) as
a function of rapidity distance from beam rapidity appears unchanged
over a wide range of beam energies \cite{PhobosSpectraAndV2,MPP,v1v4},
a pattern known as limiting fragmentation~\cite{LF}.
Figure~\ref{fig:v1Eta3Energies} presents $v_1$ results in the
projectile frame for three beam energies. In this frame, zero on the
horizontal axis corresponds to beam rapidity for each of the three
beam energies.  The data support the limiting fragmentation hypothesis
in the region $-2 < y-y_\mathrm{beam} < -1$.
  
In summary, we have presented the first measurements of charged  
particle  
directed flow in Au+Au collisions at $\sqrt{s_{_{NN}}} = 62.4$\,GeV. 
The analysis has been performed using three different methods and the 
results agree very well with each other.  One of the methods involves 
the determination of the reaction plane from the bounce-off of 
fragmentation neutrons, the first measurement of this type at RHIC. 
This method provides measurements of directed flow that are expected to  
have negligible systematic uncertainty arising from non-flow effects.   
In addition, these measurements provide a determination 
of the sign of $v_1$.  In this way, we conclude that charged particles 
in the pseudorapidity region covered by the STAR TPC and FTPCs (up to 
$|\eta|=4.0$) flow in the opposite direction to the fragmentation 
nucleons with the same sign of $\eta$.  
The $p_t$-dependence of $v_1$ saturates above $p_t \approx 
1$\,GeV/$c$ in the mid-pseudorapidity region 
and $p_t \approx 0.5$\,GeV/$c$ in the forward pseudorapidity region.   
Over the pseudorapidity range studied, 
no sign change in the slope of charged-particle $v_1$ versus pseudorapidity is
observed at any centrality.
The centrality dependence of $v_1$ in the region of $2.5 < |\eta| < 4.0$ 
is found to be stronger than what is observed closer to mid-pseudorapidity.
The rapidity dependence of $v_1$
provides further support for the limiting fragmentation 
picture. 
  
\begin{acknowledgments}          
We thank Sebastian White for consultations and help in construction of   
the STAR ZDC-SMD, and we acknowledge the RHIC Operations Group and RCF   
at BNL, and the NERSC Center at LBNL for their support. 
This work was supported
in part by the HENP Divisions of the Office of Science of the U.S.
DOE; the U.S. NSF; the BMBF of Germany; IN2P3, RA, RPL, and
EMN of France; EPSRC of the United Kingdom; FAPESP of Brazil;
the Russian Ministry of Science and Technology; the Ministry of
Education and the NNSFC of China; IRP and GA of the Czech Republic,
FOM of the Netherlands, DAE, DST, and CSIR of the Government
of India; Swiss NSF; the Polish State Committee for Scientific 
Research; STAA of Slovakia, and the Korea Sci. \& Eng. Foundation.
\end{acknowledgments}          
  
\end{document}